\documentclass[preprint,10pt]{aastex}

\slugcomment{} 

\shorttitle{The FUSE Spectrum of
PG\,0804\,+761}
\shortauthors{Richter et al.}

\begin{document}

\title{The FUSE Spectrum of
PG\,0804\,+761: A Study of Atomic and Molecular
Gas in the Lower
Galactic Halo and Beyond}


\author{Philipp Richter, Blair. D. Savage,
Bart P. Wakker}
\affil{Washburn Observatory, University of Wisconsin-Madison,
475 N. Charter Street, Madison, WI\,53706}
\email{richter@astro.wisc.edu, savage@astro.wisc.edu, wakker@astro.wisc.edu}
\author{Kenneth R. Sembach}
\affil{Department of Physics and Astronomy, Johns Hopkins University,
3400 N. Charles Street, Baltimore MD\,21218}
\email{sembach@pha.jhu.edu}
\and
\author{Peter M.W. Kalberla}
\affil{Universit\"at Bonn, Radioastronomisches Institut, D-53121 Bonn, Germany}
\email{kalb@astro.uni-bonn.de}




\begin{abstract}
We present an analysis of interstellar and intergalactic absorption
lines in the FUSE spectrum of the low-redshift quasar PG\,0804\,+761 ($z_{\rm em}=0.100$) at
intermediate resolution (FWHM$\sim 25$ km\,s$^{-1}$) in the direction
$l=138\fdg3$, $b=31\fdg0$. 
With a good signal-to-noise ratio (S/N) and the presence
of several interesting Galactic and extragalactic absorption
components along the sight line, this
spectrum provides a good
opportunity to demonstrate the ability of FUSE
to do both interstellar and extragalactic science.
Although the spectrum 
of PG\,0804\,+761 is dominated by strong absorption from
local Galactic gas at 0 km\,s$^{-1}$, 
we concentrate our
study on absorption by molecular hydrogen and neutral and ionized metals
related to an intermediate-velocity cloud (IVC)
in the lower Galactic halo at $-55$ km\,s$^{-1}$, and on absorption 
from 
O\,{\sc vi} extended to negative velocities. 
In the IVC, weak molecular hydrogen absorption is
found in 5 lines for rotational levels $0$ and $1$,
leading to a total H$_2$ column density of
log $N = 14.71 \pm 0.30$.
We derive an O\,{\sc i} gas-phase 
abundance for the IVC 
of $1.03^{+0.71}_{-0.42}$ solar. 
Lower abundances of
other elements (Fe, Si) imply depletion onto dust grains
or the presence of higher, undetected
ionization states.
The presence of N\,{\sc ii} and Fe\,{\sc iii} absorption at $-55$ km\,s$^{-1}$
indicates that a fraction of the hydrogen is ionized.
From the relative abundances of O\,{\sc i} and P\,{\sc ii} we estimate a degree of
ionization H$^+$/(H$^0 +$H$^+$) of $\sim 19$ percent.
Absorption by O\,{\sc vi} is found at
velocities as negative as $-110$ km\,s$^{-1}$,
but no absorption from any species
is found at velocities of $\sim -180$ km\,s$^{-1}$ where absorption
from the nearby high-velocity cloud Complex A would be expected.
We suggest that the extended
O\,{\sc vi} absorption traces hot gas
situated above the Perseus spiral arm. 
Finally, we find intergalactic absorption
by an intervening H\,{\sc i} Ly\,$\beta$ absorber at $z_{\rm abs}=0.019$ 
and absorption by H\,{\sc i}, C\,{\sc iii} and O\,{\sc vi} 
in an associated system 
at $z_{\rm abs}=0.102$. No intervening O\,{\sc vi} absorbers are seen in the spectrum
of PG\,0804\,+761.

\end{abstract}




\section{Introduction}

The succesful launch of the {\it Far Ultraviolet Spectroscopic Explorer} (FUSE)
in June 1999 (Moos et al.\,2000) has provided a major new
facility for obtaining 
absorption line spectra at intermediate resolution 
(FWHM$\sim 25$ km\,s$^{-1}$)
in the far ultraviolet (FUV) regime
from 905 to 1187 \AA.
In this wavelength range, 
many spectral diagnostic lines provide information over a wide range of 
physical conditions found in
the diffuse and translucent interstellar
medium. The Lyman and Werner bands of
molecular hydrogen (H$_2$), as well as absorption bands
of carbon monoxide (CO), give important information about
the cold molecular gas. The
complete Lyman series of neutral hydrogen and deuterium,
in combination with a large number of transitions of
neutral and weakly ionized metals, can be used to study the conditions
of the neutral and diffuse ionized gas in great detail.
Finally, absorption lines from highly ionized species, in particular the O\,{\sc vi}
doublet near 1030\,\AA, are excellent tracers of
the hot component of the ISM, as found in halos of
galaxies or within expanding shells and superbubbles.
Absorption spectroscopy in the FUV
range is an extremely powerful technique for the
study of the multi-phase structure of the ISM
of the Milky Way and gives additional information
about the structure and physics of the intergalactic
medium (IGM), when applied to quasars or Active Galactic Nuclei (AGNs).

FUV absorption spectroscopy with the
{\it Copernicus} satellite (1972-1981) was 
performed to intensively study the 
interstellar medium (for a review see 
Spitzer \& Jenkins 1975), but was limited in
sensitivity to
nearby bright stars as background sources.  
Later satellites, such as the {\it International
Ultraviolet Explorer} (IUE) and the
{\it Hubble Space Telescope} (HST), did not cover wavelengths
below 1150 \AA. The {\it Orbiting and Retrievable Far and
Extreme Ultraviolet Spectrometer} (ORFEUS), launched
in 1996 for a 14-day {\it Space Shuttle} mission, was 
able to obtain intermediate resolution FUV spectra
toward extragalactic background sources. Observations
with ORFEUS led to many new
insights concerning the ISM in the Galactic halo and the
Magellanic Clouds
(Widmann et al.\,1998; de\,Boer et al.\,1998; Richter et al.\,1998, 1999),
but the number of observed
objects was limited due to the short mission duration.
With FUSE, the number of extragalactic sight-line spectra
in the FUV range is growing rapidly and
the very first results from FUSE
(e.g., Oegerle et al.\,2000, Savage et al.\,2000; Sembach et al.\,2000;
Shull et al.\,2000; Snow et al.\,2000) already imply that the
satellite will make important contributions to
studies of the ISM and IGM.
With its high sensitivity and its
moderate resolution, FUSE is able to
obtain spectra toward fainter (and thus more distant)
background sources and so gives access to regions in space which could not be
studied before in the FUV range.

In this paper we present measurements of absorption 
lines in the spectrum of 
the bright low-redshift quasar
PG\,0804\,+761 ($V=15.20, z_{\rm em} =0.100$, Schmidt \& Green 1983) obtained with
FUSE in January 2000.
The line of sight to PG\,0804\,+761 in the Galactic
direction $l=138\fdg3, b=+31\fdg0$ passes through
local Galactic gas and through the Galactic halo
above nearby and distant spiral arms.
The FUSE spectrum of PG\,0804\,+761
thus provides an opportunity
to investigate the ISM along a path extending
over the outer regions of the Milky Way.
The goal of this paper is to analyze the sight-line
structure in this interesting direction
and to provide a reference
spectrum for further studies.

The outline of this paper is as follows:
a review of the
sight-line structure is given in \S2.
In \S3 we present the FUSE observations
and the data reduction. 
Absorption by local Galactic gas is briefly 
discussed in \S4.
Absorption by molecules, neutral and weakly
ionized species at intermediate negative
velocities is presented in \S5, while
absorption by moderately and highly ionized gas is
described in \S6. 
In \S7 we shortly discuss the lack of absorption by high-velocity gas 
from Complex~A. 
\S8 presents the results
for the intergalactic gas and the associated system in front
of PG\,0804\,+761. Finally, a summary
of our study is given in \S9.

\section{The Sight-line Structure
toward PG\,0804\,+761}

The absorption along the line
of sight to PG\,0804\,+761 ($l=138\fdg3, b=+31\fdg0$) is primarily
dominated by various Galactic disk and
halo gas-components. The sight line
passes through local Galactic gas,
through intermediate-velocity gas
in the Galactic halo and over the 
outer spiral arms of the Milky Way.
The position of PG\,0804\,+761 lies $< 0\fdg5$ off
the edge of high-velocity cloud
Complex A (as seen in 21\,cm; Wakker \& van Woerden 1997), 
so that weak absorption from the outermost edge of Complex A could 
be present in spectrum of PG\,0804\,+761.
In the direction $l=138\fdg3$, $b=31\fdg0$ 
differential Galactic rotation has a systematic influence
on the radial velocities.
In particular, if halo gas is co-rotating with the
underlying disk, velocities
in the range $0$ to $-55$ km\,s$^{-1}$ are predicted for gas 
with $z <5$ kpc, assuming a flat rotation curve for the
outer Galaxy and a LSR rotation speed
of $220$ km\,s$^{-1}$.
Additional intergalactic absorption features
are expected as well, given the
redshift of $z_{\rm em}= 0.100$ for PG\,0804\,+761.

In this study, we concentrate on absorption associated with the gas at
intermediate velocities around $-55$ km\,s$^{-1}$. This component is part of the
``Low-Latitude Intermediate-Velocity Arch'' (LLIV Arch) defined in H\,{\sc i} 21\,cm emission
(Kuntz \& Danly 1996). It is a $5-10\degr$ wide, elongated structure, running
from ($l$,$b$)$\sim$($115\degr$,$37\degr$) to ($l$,$b$)$\sim$($165\degr$,$35\degr$), and
having velocities in the range $-30$ to $-70$ km\,s$^{-1}$. Fig.\,1 shows an H\,{\sc i} contour map
of the gas in this velocity range, based on data in the Leiden-Dwingeloo Survey 
(Hartmann \& Burton 1997). Six cores are labeled LLIV1 to LLIV6, as
defined by Kuntz \& Danly (1996). The contours show the high-velocity gas at
$v_{\rm LSR}<-100$ km\,s$^{-1}$
in this region of the sky.
A distance bracket for the LLIV Arch of 0.9--1.8\,kpc ($z$=0.6--1.2\,kpc) is determined 
by Wakker (2000), using the detection of Ca\,{\sc ii} and Mg\,{\sc ii} absorption
at its velocity in four background horizontal branch stars, combined with
significant non-detections of Ca\,{\sc ii} and Mg\,{\sc ii} in three other stars (Vladilo et al.
1994; Welsh, Craig \& Roberts 1996; Wakker et al. 1996; Ryans et al. 1997).
In the Galactic plane the Perseus Arm spans the longitude range of the LLIV Arch.
According to Reynolds et al.\ (1995), this spiral arm lies at a
distance of $\sim$2.5\,kpc from the Sun. The LLIV Arch thus appears to lie about
1\,kpc above the plane in the inter-arm region between the Perseus and the Local
spiral arm. It deviates by about $-40$ km\,s$^{-1}$ from the velocity expected for a
co-rotating cloud at its position. 
An approximate metallicity for the LLIV Arch was previously derived from the IUE
spectrum of SN\,1993\,J ($l=142\fdg2$, $b=40\fdg9$; de\,Boer et al.\,1993; Vladilo et
al.\,1993). In this data, the LLIV Arch shows
an abundance pattern typical for warm disk gas (as defined by Savage \& Sembach
1996). As Zn\,{\sc ii} is generally undepleted, its abundance of 1.6$\pm$0.4 times solar
suggested a slightly supersolar intrinsic abundance for the LLIV Arch.

Two portions of the FUSE spectrum of PG\,0804\,+761 are presented 
in Fig.\,2.
The distribution of H\,{\sc i} 21\,cm
emission along the sight line, as 
measured with the 100\,m radio telescope in Effelsberg 
(Wakker et al.\,2000; $9\farcm1$ beam),
is shown in the upper left panel of Fig.\,3.
While the strong local Galactic component
peaks near 0 km\,s$^{-1}$, gas from the LLIV Arch shows
its emission maximum at $-57$ km\,s$^{-1}$  
with T$_{\rm B} \simeq 0.6$ K, equivalent to a column
density of $N($H\,{\sc i}$)= 3.45 \pm 0.09 \times 10^{19}$ cm$^{-2}$.
This value agrees well with other H\,{\sc i} emission line observations
toward PG\,0804\,+761 at lower angular resolution, such as obtained with the NRAO 140\,ft
(Murphy, Sembach \& Lockman, in preparation; $21 \arcmin$ beam) or the Leiden-Dwingeloo Survey
(Hartman \& Burton 1997; $36 \arcmin$ beam).
No H\,{\sc i} emission is seen at velocities more negative than
$-100$ km\,s$^{-1}$, where gas from Complex A is detected
$< 0\fdg5$ away (Wakker \& van Woerden 1997).

The line of sight ends at the bright ($V=15.2$) quasar PG\,0804+761,
in the literature cited to be at a redshift of $z_{\rm em}=0.100$ 
(Schmidt \& Green 1983). Shull et al.\,(2000) reported absorption from an
intervening Ly\,$\alpha$ absorber along this sight line
at $+5565$ km\,s$^{-1}$ or $z_{\rm abs} = 0.019$, using 
HST STIS data. The latter component is also seen in Ly\,$\beta$
absorption in the FUSE spectrum of PG\,0804+761
(Shull et al.\,2000). 
The STIS data, however, do not cover the wavelength range
above 1300 \AA, so there is no information
about intervening Ly\,$\alpha$ systems for redshifts 
$z>0.069$. We will see in \S5 that the FUSE data
reveal one more intergalactic absorption 
component, which is most likely associated with
PG\,0804+761 itself.

\section{Observations and Data Reduction}

\subsection{Observations}

FUSE observations of PG\,0804\,+761
were carried out on 1999 October 5 as one
of the very first objects observed with FUSE.
At this early state, FUSE had not reached
its optimal performance so that this early
data set was used only for a qualitative
overview of the sight-line structure.
Additional obervations, with higher resolution
and higher S/N, were performed in January 2000, 
using the large aperture (LWRS, $30\farcs0 \times 30\farcs0$) in the time-tag
(TTAG) mode, in which the X and Y locations, the arrival time, and
the pulse height of each detection are stored as a photon list.
The quantitative analysis of the line of sight
toward PG\,0804\,+761 presented in the following sections relates
to the data obtained in January.
Basic information about these observations is given
in Table 1.
FUSE consists of four coaligned prime focus
telescopes and Rowland-type spectrographs in combination
with microchannel plate detectors. Two of the
telescope channels have Al:LiF coatings, optimized for
the wavelength range from 1000 to 1187 \AA.
The two other channels use SiC coatings and are
sensitive from 905 to 1105 \AA.
More detailed descriptions of the instrument
and its on-orbit performance are presented by Moos et al.\,(2000) and
Sahnow et al.\,(2000).
The spectral resolution of the
PG\,0804\,+761 data is $\lambda / \Delta \lambda \approx 12,000$,
equivalent to $25$ km\,s$^{-1}$, as estimated from 
profile fits for some of the weaker H$_2$ absorption lines.
The average continuum flux in the spectrum is $\sim 8 \times
10^{-14}$ erg cm$^{-2}$ s$^{-1}$ \AA$^{-1}$. 

\subsection{Data Reduction and Analysis Method}

The total observing time for the
spectrum of PG\,0804\,+761 was 21,125\,sec, recorded
in five individual exposures (see Table 1). 
At the time that this paper was prepared, the
absolute wavelength
calibration for the early FUSE data was uncertain.
We therefore measured the
line centers for 20 absorption lines
of molecular hydrogen in the local Galactic
gas component and calibrated the wavelength
scale by fixing the local H$_2$ gas
component at $v_{\rm LSR} \sim 0$ km\,s$^{-1}$,
representing the average of the two strong Galactic
H\,{\sc i} emission components around zero-velocities
(Wakker et al.\,2000).
Wavelengths for the H$_2$ transitions are taken
from the list of Abgrall \& Roueff (1989); for atomic
species we use the compilation of Morton (in preparation).
From the internal scatter of the velocities
for the H$_2$ line centers we estimate
a mean accuracy of $\sim 10$ km\,s$^{-1}$ for our
wavelength calibration, but velocity deviations of more
than $20$ km\,s$^{-1}$ still remain for some absorption lines 
in the spectrum. For these cases, the velocity scale was
adjusted for each line individually with respect to the nearest 
reference line.

The data was extracted using the CALFUSE (Vers.1.6.6)
standard reduction pipeline, providing the 
complete set of the individual exposures 
in each channel. 
The single exposures were coadded for each of
the channels. In order to increase the signal-
to-noise ratio (S/N), channels covering
the same wavelength ranges were also combined.
The latter procedure was made for each line
individually, correcting for the deviations
in the wavelength scale by adjusting the
(Gaussian) line centers for the local
Galactic component (at $0$ km\,s$^{-1}$) to
the zero point of the overall wavelength 
scale, as described above.
We are aware of the fact that this combination
introduces systematic errors
in the flux distribution over the 
pixel scale. The increase of S/N,
however, justifies this procedure in view
of the reduced uncertainty for the 
determination of absorption line equivalent widths.
The spectra were rebinned over
three pixels, but no additional smoothing
was applied to the data. After combining and rebinning, 
the typical S/N turns out to be $\sim 8$ per (binned) pixel element.
Fig.\,2 shows part of the summary spectrum 
(all exposures, all channels) of 
PG\,0804\,+761, plotted in flux units versus wavelength.
The various interstellar absorption features have
been marked and identified above the spectrum.
Terrestrial emission features are also tagged.
Fig.\,3 shows a selection of continuum normalized
interstellar absorption line profiles for
PG\,0804\,+761 plotted on a LSR velocity scale.
  
Equivalent widths $W_{\lambda}$ have been
measured by fitting multi-Gaussian components
to observed line profiles. 
This method was chosen in view
of the fact that the true line profiles from
molecular hydrogen and weakly ionized species are
not resolved in the FUSE data.
For the O\,{\sc vi} absorption, whose velocity 
profile is fully resolved at $\lambda / \Delta \lambda \approx 12,000$,
we have measured $W_{\lambda}$ by a direct pixel integration 
and its column density by making use of the apparent-optical-depth
method, as described by Savage \& Sembach (1991).
Continua were normalized by fitting low-order polynomials
to the data in the vicinity of each line. Errors for the
$W_{\lambda}$, as given in Tables 2 and 4,
include contributions from photon-counting
statistics, from the fitting procedure, and from
the continuum-placement errors.

\section{Local Galactic Gas at 0 km\,s$^{-1}$}

The spectrum of PG\,0804+761 is dominated by absorption from
molecular hydrogen and metals situated in the local Galactic
ISM. 
Over 80 Galactic absorption lines are detected in the
wavelength range between 920 and 1180 \AA.
Shull et al.\,(2000a) studied H$_2$ absorption
in the Galactic gas component in the FUSE spectrum
of PG\,0804+761, finding a total H$_2$ column
density of $N($H$_2) = 6.1 \times 10^{18}$ cm$^{-2}$
and a temperature of T$_{0,1}=113$ K.
In the present study, we do not
discuss results for the local component.
74 Galactic absorption lines from neutral and molecular hydrogen and
weakly ionized metals
have been analyzed and their equivalent widths and errors can
be made available by the authors. The metal lines 
detected include: C\,{\sc ii} ($\lambda 1036.337$), 
C\,{\sc iii} ($\lambda 977.020$), 
N\,{\sc i} ($\lambda 953.415, \lambda 1134.980$),
N\,{\sc ii} ($\lambda 1083.990$),
O\,{\sc i} ($\lambda 936.630, \lambda 971.738$),
Si\,{\sc ii} ($\lambda 1020.699$), P\,{\sc ii} ($\lambda 1152.818$),
Ar\,{\sc i} ($\lambda 1048.220, \lambda 1066.660$),
Fe\,{\sc ii} ($\lambda 1112.048, \lambda 1121.975, \lambda 1125.448,
\lambda 1133.665$,\\
$\lambda 1143.226, \lambda 1144.938$), and
Fe\,{\sc iii} ($\lambda 1122.526$).

\section{Absorption from Gas in the Lower Galactic Halo}

The FUSE spectrum of PG\,0804+761 shows
absorption associated with the LLIV Arch at $-55$ km\,s$^{-1}$ 
from the lowest two rotational
states of molecular hydrogen,
from various atomic species 
(C, N, O, Si, P, Ar, Fe), 
and from the Lyman
series of neutral hydrogen.

\subsection{Molecular Hydrogen}

Weak absorption by molecular hydrogen (H$_2$) near $-55$
km\,s$^{-1}$ associated with gas in the 
LLIV Arch is seen in several lines from the
lowest 2 rotational states ($J=0$ and $1$).
H$_2$ absorption profiles for six lines
are presented in Fig.\,3 in the left and middle panels.
Equivalent widths for the detected absorption features
range between $10$ and $50$ m\AA\, for lines
with $\lambda \ge 980$ \AA. For the strongest of
the H$_2$ transitions in the wavelength range below
980 \AA, an accurate determination of the H$_2$ line strengths
was not possible because of the low signal-to-noise ratio
in the two SiC channels and because of blending with
Galactic H$_2$ absorption or atomic
absorption lines.
For $\lambda \ge 980$ \AA\, we measured equivalent
widths for 5 lines and determined upper
limits for 7 additional lines, as listed in Table 2.
To derive column densities for the individual rotational
states of H$_2$ we fitted the measurements to a
curve of growth with $b=6$ km\,s$^{-1}$ (Fig.\,4).
This $b$-value represents the best fit to the observations.
Its uncertainty ($\approx \pm$ 1 km\,s$^{-1}$) has
only a minor influence on the determination of the column
densities, since most of the lines are on the linear
part of the curve of growth for which the choice of the
$b$-value is unimportant. We obtained the H$_2$ column
densities $N$ for $J=0$ and $1$ and upper limits for the
levels $J=2,3$,
presented in Table 3. The errors for $N$
are based on the uncertainties
for the individual equivalent widths, the error of $b$,
as well as the error for the fit to the curve of growth.
Summing over the individual column densities for $J=0,1$ we
find a total logarithmic H$_2$ column densitiy of
log $N = 14.71 \pm 0.30$ for the gas
in the LLIV Arch. The error here
includes the upper limits in $N$ for possible 
contributions from $J=2,3$.
Together with $N$(H\,{\sc i}) $=3.45 \times 10^{19}$ cm$^{-2}$
from the Effelsberg 21\,cm data, the fraction of hydrogen
in molecular form is $f=2N($H$_2)/[N($H\,{\sc i}$)+2N($H$_2)]=
2.97 \times 10^{-5}$, a value which is similar to
fractions found in local Galactic disk gas in sight lines
with low interstellar reddening (Savage et al.\,1977).
The value indicates that the observed cloud is predominantly
atomic.  
From the column densities of $J=0$ and $1$ we
find a excitation temperature of T$_{0,1}=193^{+322}_{-75}$ K by
fitting the level populations to a theoretical Boltzmann
distribution.

The detection of molecular hydrogen in the LLIV Arch
gives important information about the physical
conditions in the intermediate-velocity gas in
the lower Galactic halo. First, its presence
shows that dust is
present, since the efficient formation of 
H$_2$ requires dust grains as catalytic
reaction partners (e.g., Pirronello et al.\,1999). We shall see in the next
section that this conclusion is in agreement
with the actual metal abundance and depletion pattern
in the LLIV Arch, as derived by the analysis
of the metal lines in the FUSE spectrum.
Second, the H$_2$ excitation temperature 
(T$_{0,1}=193^{+322}_{-75}$ K) seems to be below 
the temperature expected directly after
molecule formation ($\sim 300$ K; e.g., Burton, Hollenbach \& Tielens 1992),
indicating that the H$_2$ gas has been
thermalized to a temperature of $193$ K.
Thus, the excitation temperature might
represent the kinetic temperature in
the H$_2$ gas, suggesting the existence of
cool (predominantly atomic) gas clumps in
the LLIV Arch. 
However, in view of the large uncertainties for T$_{0,1}$
and the possibility, that the excitation
is caused by thermalization processes on the grain surface,
the causal connection between the
excitation of the H$_2$ gas and the
kinetic gas temperature remains
speculative.

Molecular hydrogen in intermediate-velocity gas
has also been detected toward HD\,93521 
(Gringel et al.\,2000) and toward the LMC 
(de\,Boer 2000, priv.\,comm.). Though the molecular
fractions are generally low, 
these results underline that the conditions
in the lower Galactic halo allow molecular 
hydrogen and dust to exist, 
an aspect which certainly has to be
considered for theoretical models of 
the Milky Way halo.

\subsection{Neutral and Weakly Ionized Metals}

Absorption at $-55$ km\,s$^{-1}$ in low
and intermediate ionization states of heavy elements is seen in the
absorption profiles of
C\,{\sc ii}, N\,{\sc i}, N\,{\sc ii}, O\,{\sc i},
Si\,{\sc ii}, P\,{\sc ii}, Ar\,{\sc i}, Fe\,{\sc ii}
and Fe\,{\sc iii}.
Higher ionization states for these elements,
such as C\,{\sc iii} and O\,{\sc vi}, also
clearly show absorption extended to negative
velocities, but no separation between different
velocity components can be made with the present 
data. Absorption from these higher ions will 
be discussed in \S6.
Fig.\,3 presents normalized absorption profiles
for some of the analyzed lines. The absorption
pattern in metals is similar to that seen in 
neutral hydrogen emission (upper left panel of Fig.\,3) and
molecular hydrogen absorption. Equivalent widths, $W_{\lambda}$,
for 18 lines at $-55$ km\,s$^{-1}$ and their errors
are given in Table 4. Typical relative errors for the
equivalent widths are less than 20 percent.
A set of four Fe\,{\sc ii} lines accurately defines a curve of
growth with $b=8.0 \pm 0.7$ km\,s$^{-1}$ (Fig.\,4). This rather
high value for $b$ suggests that the velocity dispersion is primarily
determined by non-thermal motions rather than by
the true thermal widths of the lines. 
The fact that the $b$-value is higher for the ions than
for the molecular hydrogen (see previous section) suggests
that the weakly ionized material is more extended than
the molecular gas, the latter most likely being situated in
one or several confined regions.  
Column densities for the species listed above were
determined by fitting the data values of log($W_{\lambda}
/{\lambda}$) to the curve of growth with $b=8$ km\,s$^{-1}$.
The results are listed in Table\,5.  
The column density errors 
include uncertainties from the individual errors of $W_{\lambda}$,
the error in $b$, as derived from the dispersion
of the Fe\,{\sc ii} data points, 
and the uncertainty for the fit to the curve of
growth. 

\subsubsection{Ionization Effects} 

The presence of absorption by N\,{\sc ii} and
Fe\,{\sc iii} at $-55$ km\,s$^{-1}$ implies that
a significant fraction of the hydrogen is ionized.
We derive values of $-0.86$ for log($N$(N\,{\sc ii})/$N$(N\,{\sc i}))
and $-1.41$ for log($N$(Fe\,{\sc iii})/$N$(Fe\,{\sc ii})).
This suggests that an ionization correction 
is required to
determine accurate gas phase abundances in the gas of the LLIV Arch.

One can estimate the fraction
of ionized hydrogen by comparing the relative abundances
of P\,{\sc ii} and O\,{\sc i}, assuming that
the observed ionization states arise in the
same physical region (i.e., in the same IVC).
Ionization of O should be the
same as for H, since both elements have a similar
ionization potential and they are strongly coupled
through charge exchange reactions.
Phosphorus is not depleted in
dust grains (Jenkins, Savage \& Spitzer 1986) and the ionization potential
of P\,{\sc ii} is 19.73\,eV. We expect
P\,{\sc ii} to be the dominant ionization state for phosphorus,
and P\,{\sc ii} will exist where
H and O are ionized.
If we assume that 
$N$(P)/$N$(O)$_{\rm LLIV Arch} \approx$ 
$N$(P)/$N$(O)$_{\rm solar}$ we can derive
the ionization of O (and thus of H)
by scaling the observed ($N$(P\,{\sc ii})/$N$(O\,{\sc i})) ratio
in the LLIV Arch to the expected solar ratio. For the LLIV Arch we find
log($N$(P\,{\sc ii})/$N$(O\,{\sc i}))$ = -3.21$, from which
we estimate a degree of ionization H$^{+}$/(H$^{0}$+H$^{+}$)
of $\sim 19$ percent. 
This value fits well to the ionization fraction of
N\,{\sc ii} (N$^{+}$/(N$^{0}$+N$^{+}$)$=0.12$), which
should roughly scale with the hydrogen photoionization
fraction (Jenkins et al.\,2000) and thus is an independent
measure for the fraction of ionized hydrogen.
In the following section, we will use the value of 
H$^{+}$/(H$^{0}$+H$^{+}$)$=0.19$ for
the determination of gas-phase abundances in
the LLIV Arch.

\subsubsection{Gas-Phase Abundances}

For the analysis of gas phase abundances 
we use the compilation of solar reference
abundances by Anders \& Grevesse (1989), which are
primarily based on meteoric data, except for 
C, N, and O, for which abundances from 
solar photosphere data were adopted (Grevesse \& Noels 1993). 
For the H\,{\sc i} column density in the LLIV Arch we take
the value of $3.45 \pm 0.09 \times 10^{19}$ cm$^{-2}$ from
the Effelsberg data. Taking into account the ionization fraction
(see previous section),
the total hydrogen column density
is $N$(H)$=N$(H\,{\sc i})$+N$(H\,{\sc ii})
$=4.26 \times 10^{19}$ cm$^{-2}$.
The amount of hydrogen in molecular
form (\S5.1) does not contribute 
significantly to the total hydrogen column density.

The abundance
pattern for the LLIV Arch along our sight line
(Table 6, Fig.\,5) is similar to that
for diffuse warm clouds in the disk of the Milky
Way (Savage \& Sembach 1996). 
Because of the reasons described in the previous section, the O\,{\sc i}/H\,{\sc i} ratio
should be representative of the actual oxygen abundance in the
gas and thus provide a good measure for the overall
metallicity.
We find log($N$(O\,{\sc i})/$N$(H\,{\sc i}))$=-3.12 \pm 0.23$, which
is $1.03^{+0.71}_{-0.42}$ solar.
Note that there are additional systematic uncertainties
in this and the other element abundances due to the
fact that we compare UV absorption line data with
beam smeared H\,{\sc i} emission line measurements.
Another problem is that the absorption lines for O\,{\sc i} as well as the
ones from N\,{\sc i} are saturated and are located on
the intermediate part of the curve of growth where
small errors for $W_{\lambda}$ cause large
uncertainties in the determination of the column
densities.
Thus, the column densities for N\,{\sc i}, N\,{\sc ii}  and
O\,{\sc i} are most uncertain, while the
equivalent widths of their lines have the smallest relative
errors (see error bars in Fig.\,5).
The relatively low abundance of N (=N\,{\sc i}$+$N\,{\sc ii})
in comparison to O\,{\sc i} and P\,{\sc ii} (see Fig.\,5) might  
be a result of the uncertainty of $N$(N\,{\sc i}) derived
with the curve-of-growth technique.
We also note
that an additional systematic uncertainty may arise from the errors
for the oscillator strengths $f$. 
Within the error bars, however, a nearly solar
gas phase abundance of the LLIV Arch is consistent
with the individual
abundances of N, O, and P.

In contrast to N, O\,{\sc i}, and P\,{\sc ii}, the 
normalized logarithmic gas-phase abundances of  Fe\,{\sc ii} and 
Si\,{\sc ii} are $\leq -0.40$ dex (Fig.\,5), 
most likely as a result of depletion into 
dust grains. Such depletion is consistent with
the detection of H$_2$ in the LLIV Arch (see \S5.1), 
which is an independent tracer for the
presence of dust in this component. 
The strong deficiency of Ar\,{\sc i} 
([Ar\,{\sc i}/H\,{\sc i}]$= -0.68 \pm 0.07$) is most likely due to
an over-ionization of Ar\,{\sc i}.
Sofia \& Jenkins (1998) argued that
a deficiency in the Ar\,{\sc i} abundance can be explained
by its very large photoionization cross section.
They find that the Ar\,{\sc i}/H\,{\sc i} ratio
is lowered in a range between $-0.11$ and $-0.96$ dex in
regions that are significantly ionized, which is in
agreement with our measurements.
In principle, one might verify this result by 
inspecting the Ar\,{\sc ii} line at 919.781 \AA,
but the low oscillator strength and the low
S/N at these wavelengths make this line 
undetectable in our spectrum.

The nearly solar gas-phase abundances for O, N and P
in combination with the presence of molecules
and dust strongly indicates that the gas
in the LLIV Arch has its origin in the Galactic disk.
We therefore conclude that the LLIV Arch is most likely
part of the return flow of a Galactic fountain 
(Shapiro \& Field 1976; Houck \& Bregman 1990) or other
mechanism that links star formation in the disk to the
interstellar material in the lower Galactic halo.

\section{Moderately and Highly Ionized Gas}

In addition to the absorption lines of molecular hydrogen and
weakly ionized metals, the FUV range gives access to information
about moderately and highly ionized gas through the
strong absorption lines of C\,{\sc iii} at 977.020 \AA\
and the O\,{\sc vi} doublet at 1031.926 and 1037.617 \AA.
The ionization energy of C\,{\sc iii} is 47.89 eV; thus C\,{\sc iii} traces
moderately ionized gas at temperatures of $\sim 10^{4}$ K.
The C\,{\sc iii} line has a large oscillator strength ($f=0.762$)
and becomes quickly saturated because
of the high abundance of carbon in the ISM.
Strong absorption from intermediate ionization states from other species 
(e.g., Si\,{\sc iii} and Si\,{\sc iv}) has
been found in sight lines to extragalactic background
sources using STIS and GHRS data (Savage, Sembach \& Lu 1997; Fabian et al., in preparation),
suggesting
that moderately ionized gas is a dominant
consituent in the lower Galactic halo.
Unfortunately, the C\,{\sc iii} absorption at negative velocities 
is blended by the absorption profile
of the O\,{\sc i} line at $\lambda = 976.448$ \AA\,
(see Fig.\,3), thus
making an analysis of the velocity distribution 
of C\,{\sc iii} toward PG\,0804+761 impractical.

From the two O\,{\sc vi} absorption lines in the 
spectrum of PG\,0804+761 only the one at 
1031.926 \AA\, is unblended and can
be used to study the highly ionized gas (T$\simeq 3 \times 10^5$ K) 
in direction of PG\,0804+761.
Savage et al.\,(2000) included the early
PG\,0804+761 observations in their determination
of a O\,{\sc vi} scale height of the Milky Way halo.
Using data from the observations performed in November,
they found an equivalent width of $286 \pm 19$ m\AA\ for the
O\,{\sc vi} line at $1031.926$ \AA\ and derived from that
a column density of log $N$(O\,{\sc vi})$=14.50 \pm 0.04$.
With the new data set
from the observations in January we find $W_{\lambda}
=260 \pm 17$ m\AA\ and log $N$(O\,{\sc vi})$=14.38 \pm 0.06$,
integrating the O\,{\sc vi} absorption profile in the range 
between $-150$ to $+100$ km\,s$^{-1}$.
The O\,{\sc vi} $\lambda 1031.926$ line profile 
is presented in Fig.\,3. Interestingly, it shows
O\,{\sc vi} absorption extended to 
velocities as negative as $-110$ km\,s$^{-1}$.
In general, O\,{\sc vi} absorption around zero velocities
is thought to trace the thick O\,{\sc vi} halo
of the Milky Way and is seen in many extragalactic 
sightlines in different directions (Savage et al.\,2000).
High velocity O\,{\sc vi} ($|v_{\rm LSR}| > 100$ km\,s$^{-1}$)
is seen in several extended sight lines
toward AGNs and quasars (Sembach et al.\,2000).
These velocities lie beyond the integration range
expected for O\,{\sc vi} absorption in the
environment of the thick disk and the lower
Galactic halo. In all cases, high velocity O\,{\sc vi} absorption
can be related to high-velocity clouds
or known structures in the outer Galaxy 
(see Sembach et al.\,2000).

For the sight-line toward PG\,0804+761 there
are different possible explanations for
the O\,{\sc vi} absorption extending to
$-110$ km\,s$^{-1}$:

\noindent
(1) The extended O\,{\sc vi} absorption might be due
to a hot gas layer surrounding Complex A. While one
might expect that Complex A has a gaseous envelope
containing highly ionized species 
so close to its H\,{\sc i} boundaries,
we reject this possibility because H\,{\sc i}
gas from Complex A close to our sight line is seen
at $-180$ km\,s$^{-1}$ (Wakker \& van\,Woerden 1997), which is
a radial velocity $70$ km\,s$^{-1}$ more negative than
the O\,{\sc vi} absorption.\\
(2) The extended O\,{\sc vi} absorption along the
sight line might represent highly ionized gas related to the
envelope of the LLIV Arch. If there is a dynamical interaction
between the cooler intermediate-velocity gas and a
hot surrounding medium one might expect that
the radial velocities for the O\,{\sc vi} absorption
should correspond to the radial velocity of the
weakly ionized gas. Since thermal and turbulent broadening 
would smear the O\,{\sc vi} absorption profile as far
as the observed $-110$ km\,s$^{-1}$, the observed profile 
is consistent with O\,{\sc vi} absorption in the LLIV Arch.\\
(3) The extended O\,{\sc vi} absorption might be related
to hot gas above the Perseus arm.
Savage, Sembach \& Lu (1995) found C\,{\sc iv} absorption 
at $-70$ km\,s$^{-1}$ toward H1821+643 ($l=94\fdg0, b=27\fdg4$) and concluded
that this component is related to hot gas situated
1.5 kpc above the Perseus arm. 
FUSE observations of H\,1821+643 (Oegerle et al.\,2000) have also
revealed O\,{\sc vi} absorption that is likely associated
with the Perseus arm.
Several arguments favor
a similar explanation for the extended
O\,{\sc vi} absorption toward PG\,0804+761:
First, the sight line toward PG\,0804+761 passes the
Perseus arm exactly above a region where a Galactic
chimney has been found, driven by the open cluster
OC1352 in the Cassiopeia OB6 association region
at $l=134\fdg8, b=0\fdg9$ (Normandeau, Taylor \& Dewdney 1996).
If chimneys such as this are common features in
the active regions in the Milky Way's spiral arms, they
should contribute significantly to the mass transfer of hot
gas into the Galactic halo (Shapiro \& Benjamin 1991; Norman
\& Ikeuchi 1989)
and would explain the
presence of hot gas above the
Perseus arm and other active regions.
Second, due to Galactic rotation, gas associated
with the Perseus arm in this direction is seen at
velocities around $-50$ km\,s$^{-1}$
(Kepner 1970), consistent with the observed
O\,{\sc vi} absorption extended to negative 
velocities when the effects of line broadening
and instrumental bluring are included. 

Although we favor (3) for explaining the observed
O\,{\sc vi} absorption profile toward PG\,0804+761,
we cannot exclude that high-velocity O\,{\sc vi} absorption
is due to a highly ionized boundary of the LLIV Arch (2),
or a combination of both possible contributors.

\section{Complex A}

Considering that PG\,0804+761 falls just outside the known neutral edge of
Complex~A (see Fig.\,1), we might expect to see absorption associated with an ionized shell
around Complex~A.  No clear H\,{\sc i} absorption is seen in the higher Lyman
lines at Complex A velocities ($\sim -180$ km\,s$^{-1}$), thus ruling out a low density
neutral envelope
around Complex~A down to a $3\sigma$ level of log $N$(H\,{\sc i})$<14.84$.
Unfortunately, the most sensitive tracer for moderately ionized gas, the 
C\,{\sc iii} line at 977.020 \AA, is blended by Galactic O\,{\sc i} absorption
(see previous section). The N\,{\sc iii} line at 989.799 \AA\, is 
not strong enough to obtain useful information about its abundance 
at the given S/N in that wavelength region.
O\,{\sc vi} absorption extends
to velocities of $\sim -110$ km\,s$^{-1}$, but not beyond (see previous section). 
We conclude that no meaningful information
about abundances in Complex~A is yielded by the FUSE data of PG\,0804+761.
However, the observations do imply that the boundary of Complex A in this direction contains 
very little H\,{\sc i}, and O\,{\sc vi}, thus indicating that Complex~A
has a well defined, sharp edge in this direction.

\section{Intergalactic Gas}

Six features in the spectrum can not be associated with
lines from gas in the inner or outer Milky Way,
but have been identified as
intergalactic absorption lines. They are listed in Table 7.

\subsection{Intervening Absorption at $z=0.019$}

Shull et al.\,(2000b) identified weak Ly\,$\beta$ absorption
at $+5565$ km\,s$^{-1}$ ($z=0.019$) in the early FUSE spectrum of
PG\,0804+761 in a FUSE mini-survey on intergalactic Ly\,$\beta$ 
absorbers at low redshifts. At this velocity, absorption by Ly\,$\alpha$
was previously detected in the STIS spectrum of PG\,0804+761
(see Shull et al.\,2000b). They give a restframe equivalent width
of $W_{\lambda}=90 \pm 30$ mA for the Ly\,$\beta$ line, based on 
the early FUSE data from November 1999. With the new data, we find Ly\,$\beta$ absorption
at 1044.710 \AA\,(equivalent to $+5553$ km\,s$^{-1}$ or $z=0.0185 \pm 0.0004$) and
$W_{\lambda}$(rest)$ =73 \pm 22$ m\AA,
in agreement with the earlier results. No O\,{\sc vi} absorption is 
seen at $z=0.0185$. $3 \sigma$ upper limits for restframe equivalent widths of the two
O\,{\sc vi} absorption lines at $\lambda_{\rm rest} = 1031.926$ and $1037.617$ \AA\,
are 64 and 52 m\AA, respectively.

\subsection{The Associated System at $z=0.102$}

The strongest intergalactic absorption component is found
at $z=0.102$. This feature is most likely related to the quasar PG\,0804+761
itself (an ``associated system'' or ``intrinsic absorber'').
This absorber is not seen in the HST data since the
wavelength coverage of the STIS spectrum does not extend above 1300 \AA,
equivalent to a limiting redshift of $z=0.069$ for the Ly\,$\alpha$
line.
The associated system is seen in neutral hydrogen (Ly\,$\beta$,
Ly\,$\gamma$) as well as in moderately (C\,{\sc iii}) to
highly (O\,{\sc vi}) ionized species. 
Normalized line profile
plots on a $z_{\rm abs} = 0.102$ restframe velocity scale for
the detected lines (except C\,{\sc iii} which is blended by
Galactic H$_2$ R(0),2-0) 
are presented in Fig.\,6. 
Restframe equivalent widths and column densities are given in Table 7.
Based on the uncertainty of the 
wavelength calibration, we estimate an uncertainty for the
redshift seen in absorption of $\sigma (z) \approx 0.0004$.

Interestingly, we see the associated system at $z_{\rm abs}=0.102$,
while PG\,0804+761 is cited in the literature to be at $z_{\rm em}=0.100$,
based on the Bright Quasar Survey from Schmidt \& Green (1983),
a subsample of the Palomar-Green Survey (Green, Schmidt \& Liebert 1986).
Their redshift is based on the measurement of two O\,{\sc iii} emission lines at 
redshifted wavelengths of $\lambda 5453$ and $\lambda 5507$ 
at a resolution of 10\,\AA (Green, Schmidt \& Liebert 1986). 
The redshift displacement between their emission line measurements
and our absorption line analysis (at $\sim 0.1$ \AA\, resolution) is $\Delta z = 0.002$,
which is similar to one resolution bin in their data and thus roughly representative
of the uncertainty in their cited redshifts. From the estimated
error of $\sigma (z) \approx 0.0004$ in the FUSE data, it is therefore probable
that the actual redshift of PG\,0804+761 is $z=0.102$ rather
than $z=0.100$. 
It cannot be excluded, however, that the offset between $z_{\rm em}$ and
$z_{\rm abs}$ is real and thus reflects the intrinsic motion
of the absorber with respect to the quasar.

The associated absorber clearly shows substructure in 
its line profiles, seen in H\,{\sc i} as well as in
the stronger one of the two O\,{\sc vi} lines (see Fig.\,6).
We identify at least two components, the stronger one at
a restframe velocity of $0$ km\,s$^{-1}$, and a weaker
component near $-45$ km\,s$^{-1}$. More absorption components
might be present but are not resolved in the FUSE data.
The absorption in the associated system could be due to 
gas in and/or around the host galaxy of PG\,0804\,+761 itself. 
Another possibility is that this gas is related
to neutral and highly ionized gas in a different system
in front of PG\,0804\,+761, possibly 
located in a galaxy cluster together with PG\,0804\,+761.
The logarithmic column densities of H\,{\sc i} and O\,{\sc vi} 
are $\sim 14.3$ and $\sim 14.0$, respectively (see Table 7). 
They are close to values
found for intervening O\,{\sc vi} absorbers, such as the system
at $z=0.142$ toward PG\,0953+415 (Tripp \& Savage 2000).
It is not clear whether low redshift associated and intervening systems
share the same physical conditions. While collisional ionization
is favoured for the intervening O\,{\sc vi} absorber toward PG\,0953+415 (Tripp \& Savage 2000),
the associated system toward PG\,0804\,+761 could also be photoionized
by the quasar itself or other surrounding photon sources in
combination with the overall EUV background field.
Additional absorption line measurements with higher resolution in
combination with optical data
are required to study structure, abundances and ionization conditions in the
associated system toward PG\,0804\,+761 in more detail.

\subsection{Intervening O\,{\sc vi} Systems}

In the following, we briefly discuss the fact that 
we do not find any intervening O\,{\sc vi}
absorption systems along a sight line with a redshift path of 0.102 and 
its implications for
the estimate of the number density per unit redshift of
intergalactic O\,{\sc vi} absorbers, recently discussed
in the work by Tripp, Savage \& Jenkins (2000).
These authors give a value of $dN/dz > 17$ for absorption
with $W_{\lambda} \ge 30$ m\AA\, at a 90 percent
confidence level for these reservoirs
of hot baryonic matter in intergalactic space, based
on HST STIS data toward the quasar H1821\,+643 ($z=0.297$).
While the redshift 
of PG\,0804+761 is $0.102$, the actual redshift range
which can be investigated for O\,{\sc vi} absorbers is
much smaller due to blending with the various ISM absorption features.
We separate intervening from associated systems at a velocity
of 6000 km\,s$^{-1}$ away from the quasar, corresponding to 
a limiting wavelength of 1121 \AA. 
All identified ISM absorption features in the spectrum 
between 1032 and 1121 \AA\, obscure $\sim 16.5$ \AA,
as measured by integrating their individual 
blocking in the spectrum. 
Since we require the detection of {\it both} lines of the O\,{\sc vi} doublet
to provide a definite detection, 
the effective blockage is $\sim 33$ \AA\, of a total of 89 \AA.
Thus, the redshift path in the spectrum of PG\,0804+761 suitable 
to investigate intervening O\,{\sc vi} systems is 
reduced from $0.102$ to $0.054$.
Assuming that $dN/dz \approx 17$ for these systems
with $W_{\lambda} \ge 30$ m\AA, one would expect
one intervening O\,{\sc vi} system toward PG\,0804+761.
The average detection limit in the FUSE spectrum of PG\,0804+761, however,
is $\sim 50$ m\AA, so that the expected number of O\,{\sc vi} systems
visible in the FUSE spectrum will be substantially smaller than one.
Therefore, the non-detection of intervening O\,{\sc vi} systems
in this data set is consistent with expectations based on the small 
unobscured redshift path and the relatively large limiting 
equivalent width.

\section{Summary}

We present FUSE far ultraviolet observations at
intermediate resolution of interstellar and intergalactic
absorption lines along the line of sight toward the
quasar PG\,0804\,+761 at $z_{\rm abs}=0.102$.
Along this sight line we find absorption from molecules and weakly 
ionized species in local
Galactic gas at $0$ km\,s$^{-1}$, and in gas associated with the 
Low Latitude Intermediate Velocity Arch (LLIV Arch) in
the lower Galactic halo at
$-55$ km\,s$^{-1}$. We also detect absorption from highly ionized
gas in the Galactic halo extended to velocities as negative
as $-110$ km\,s$^{-1}$, and from intergalactic
gas at redshifts of $z=0.019$ and $0.102$,
the latter most likely being associated with the quasar itself.

1. We find H$_2$ absorption in gas of the LLIV Arch with
a total column density of log $N=14.71 \pm 0.3$
from the lowest 2 rotational states ($J=0$ and $1$).
An excitation temperature of $193^{+322}_{-75}$ K is derived for the
H$_2$ gas from a Boltzmann fit to the relative populations
of the rotational states $J=0,1$. From the presence
of H$_2$ in the LLIV Arch we conclude that dust is 
present in this cloud in the lower Galactic halo.

2. An analysis of metal abundances in the LLIV Arch
is performed, including absorption lines from 
C\,{\sc ii}, N\,{\sc i}, N\,{\sc ii}, O\,{\sc i}, Si\,{\sc ii},
P\,{\sc ii}, Ar\,{\sc i}, Fe\,{\sc ii} and Fe\,{\sc iii}.
For O\,{\sc i}, we find a gas phase
abundance of $1.03^{+0.71}_{-0.42}$ solar. 
The presence of N\,{\sc ii} and
Fe\,{\sc iii} indicates that a substantial fraction of the
gas is ionized.
We estimate a degree of ionization
of H$^+/$(H$^0+$H$^+$) $\approx 0.19$.
The presence of dust, molecules and
metals at nearly solar abundances lead us to conclude that
the LLIV Arch is part of the return flow of a Galactic fountain.

3. We investigate velocity profiles of the
moderately and highly ionized species 
C\,{\sc iii} and O\,{\sc vi}. While C\,{\sc iii}
absorption is blended by an O\,{\sc i} line,
we find O\,{\sc vi} absorption at $\lambda = 1031.926$ \AA\,
extended to negative
velocities as far as $-110$ km\,s$^{-1}$.  We suggest
that the O\,{\sc vi} absorption at these negative velocities is related
to highly ionized gas situated above the Perseus spiral arm. Another 
possibility is that the O\,{\sc vi} absorption is due to a highly 
ionized envelope of the LLIV Arch.
           
4. Six intergalactic absorption lines of
H\,{\sc i}, C\,{\sc iii} and O\,{\sc vi} 
are found at redshifts of $z=0.019$ 
and $0.102$. The latter absorber is probably associated with 
the quasar PG\,0804+761 itself. The non-detection of 
intervening O\,{\sc vi} absorbers adds information
about the number density per unit redshift of
intergalactic O\,{\sc vi} absorbers at low redshift.

\acknowledgments

We thank the FUSE Science and Operations Teams for their 
dedicated efforts to make the observations described in this 
paper possible. We also thank Todd M. Tripp for helpful
comments on the IGM part of this study.
This work is based on data obtained for the
the Guaranteed Time Team by the NASA-CNES-CSA FUSE 
mission operated by the Johns Hopkins University. 
Financial support has been provided by NASA
contract NAS5-32985.

\clearpage
\resizebox{1.00\hsize}{!}{\includegraphics{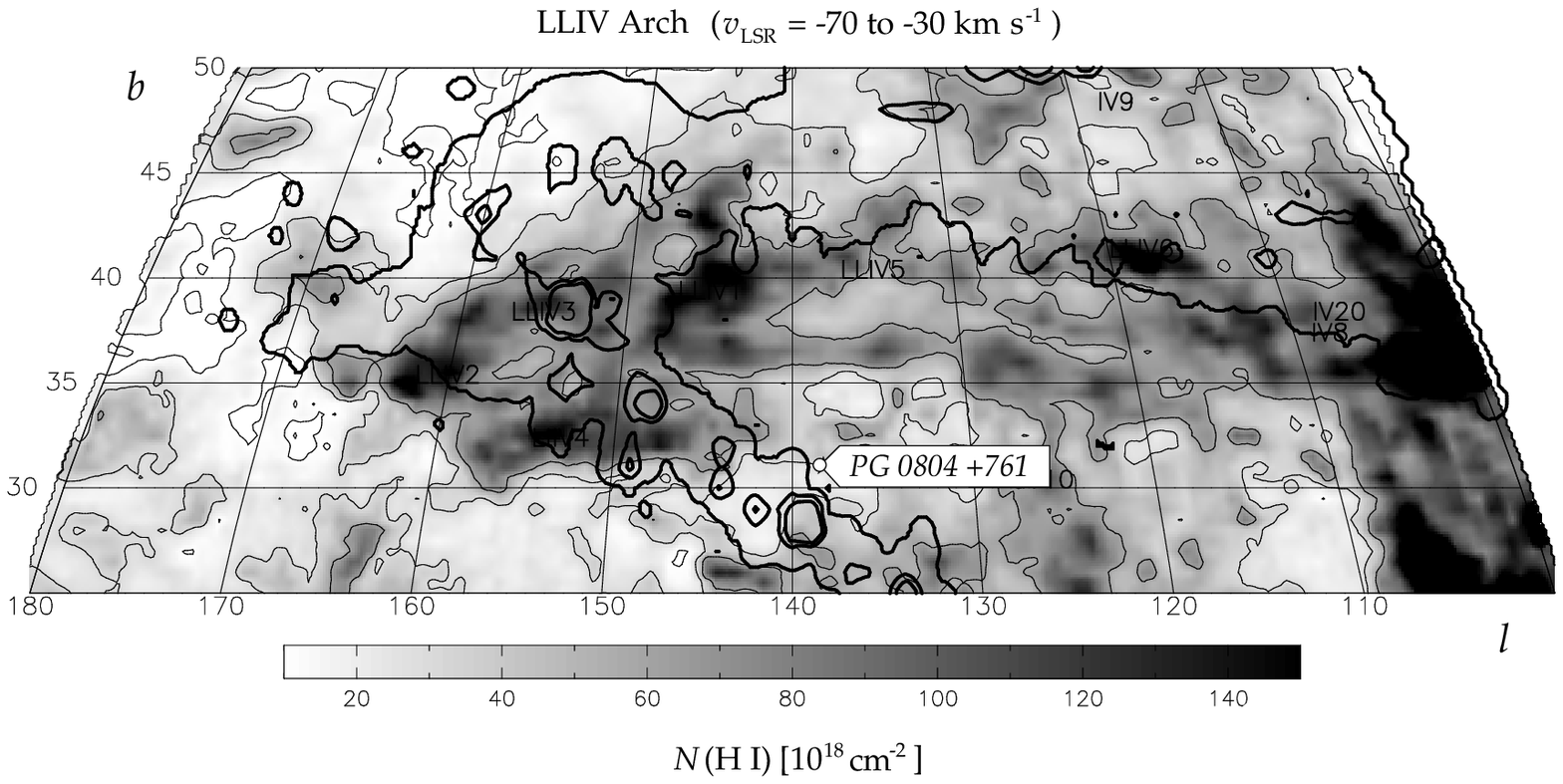}}
\figcaption[pg804fig01.ps]{
The greyscale and thin contours show the H\,{\sc i} column density of the
LLIV Arch (Kuntz \& Danly 1996), in the velocity range between $-70$ and $-30$ km\,s$^{-1}$, from the Leiden-Dwingeloo Survey
(Hartmann \& Burton 1997). Contour levels are at 1, 4, 8 and
16$\times 10^{19}$ cm$^{-2}$.
The thick contours give brightness temperature levels of
0.05, 0.7 and 1.2\,K for the high-velocity gas ($v_{\rm LSR}$$<$$-$100\,km\,s$^{-1}$)
from the data of Hulsbosch \& Wakker (1988). These approximately correspond to
column densities of 0.25, 3 and 5 $\times 10^{19}$ cm$^{-2}$. 
\label{fig1}}

\clearpage
\resizebox{0.7\hsize}{!}{\includegraphics{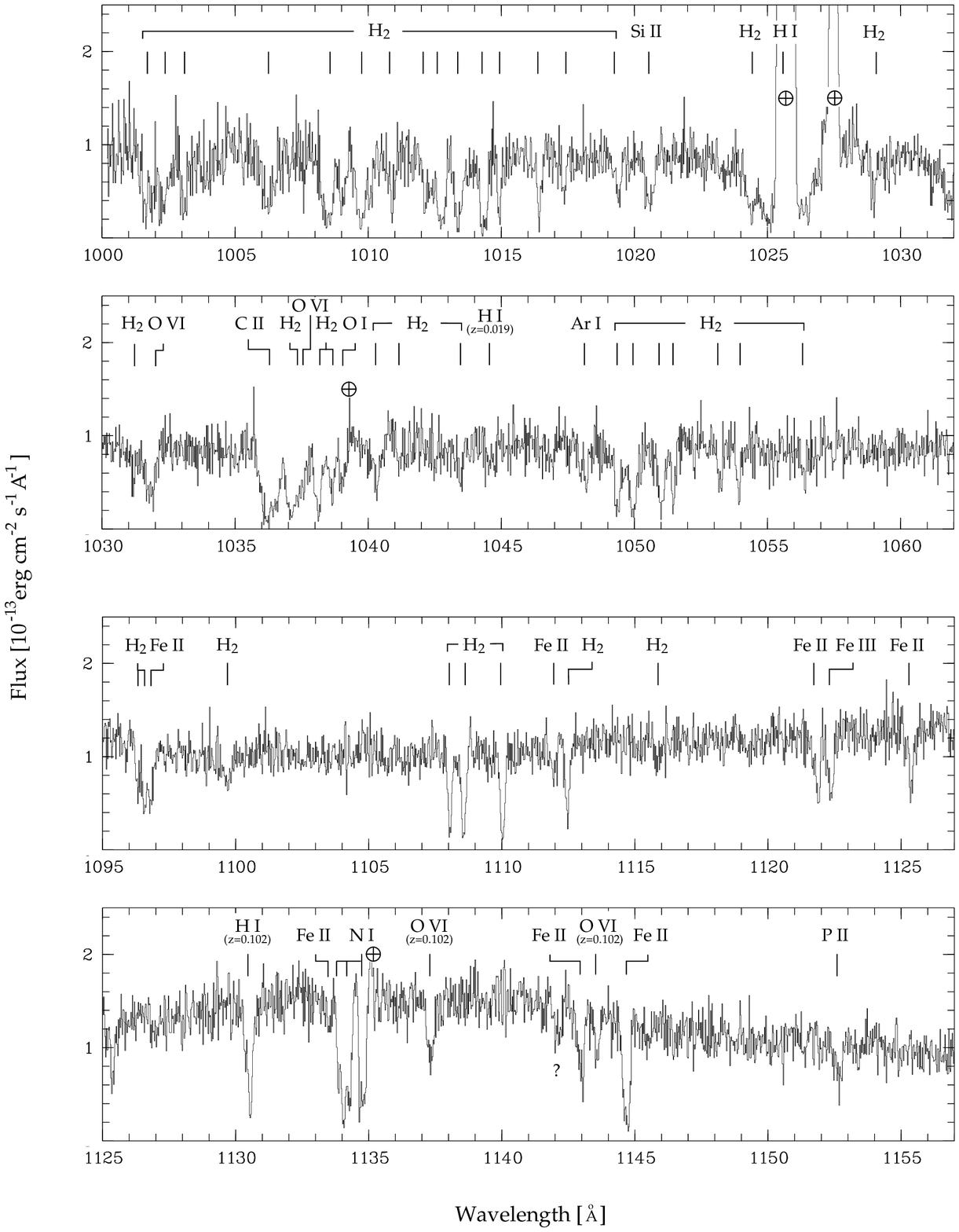}}
\figcaption[pg804fig02.ps]{
Portions of the FUSE spectrum of PG\,0804\,+761 over the wavelength ranges from 
$1000-1062$ \AA\, and $1095-1157$ \AA\, are shown. 
The plotted data represent a summary spectrum combining all channels and their segments.
The data has been binned over 3 pixels.
Galactic molecular and atomic absorption
lines (centered at their local Galactic component at $v_{\rm LSR} = 0$ km\,s$^{-1}$) and
intergalactic absorption features (presented together with their redshifts) are 
identified above the spectrum.
Terrestrial emission features are marked with a $\oplus$.
\label{fig2}}

\clearpage
\resizebox{0.7\hsize}{!}{\includegraphics{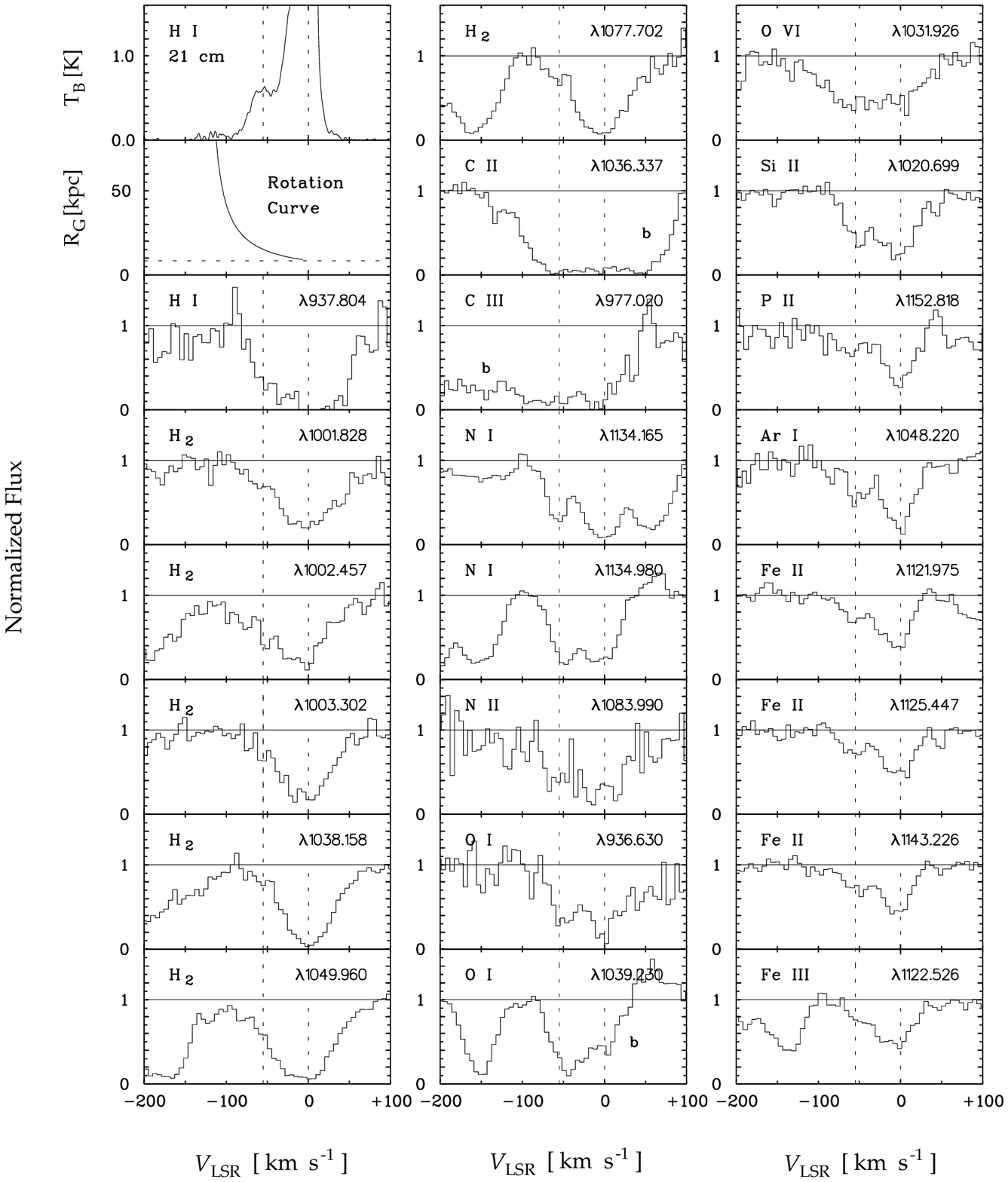}}
\figcaption[pg804fig03.ps]{
Continuum normalized interstellar absorption profiles for PG\,0804\,+761 are
plotted against LSR velocity. They include absorption from molecular
hydrogen and atomic species at low and high ionization states. The individual 
profiles are labeled 
within each box.
Local Galactic absorption is seen at $0$ km\,s$^{-1}$; absorption
associated with the LLIV Arch appears at $-55$ km\,s$^{-1}$. Both velocities
are marked with dotted lines. The O\,{\sc vi} absorption line (upper right box)
shows absorption at velocities as negative as $-110$ km\,s$^{-1}$.
At the top of the left panel we show the Effelsberg H\,{\sc i} 21\,cm emission temperature profile
(Wakker et al.\,2000; $9\farcm1$ beam)
and the relation between LSR velocity and Galactocentric distance,
$R_{\rm G}$, for co-rotating gas and a flat Galactic rotation curve 
with $\Theta$($R_{\rm G} \ge 8.5$ kpc)$=220$ km\,s$^{-1}$. 
Regions blended by other lines 
are labeled with ``b''.
\label{fig3}}

\clearpage
\resizebox{0.85\hsize}{!}{\includegraphics{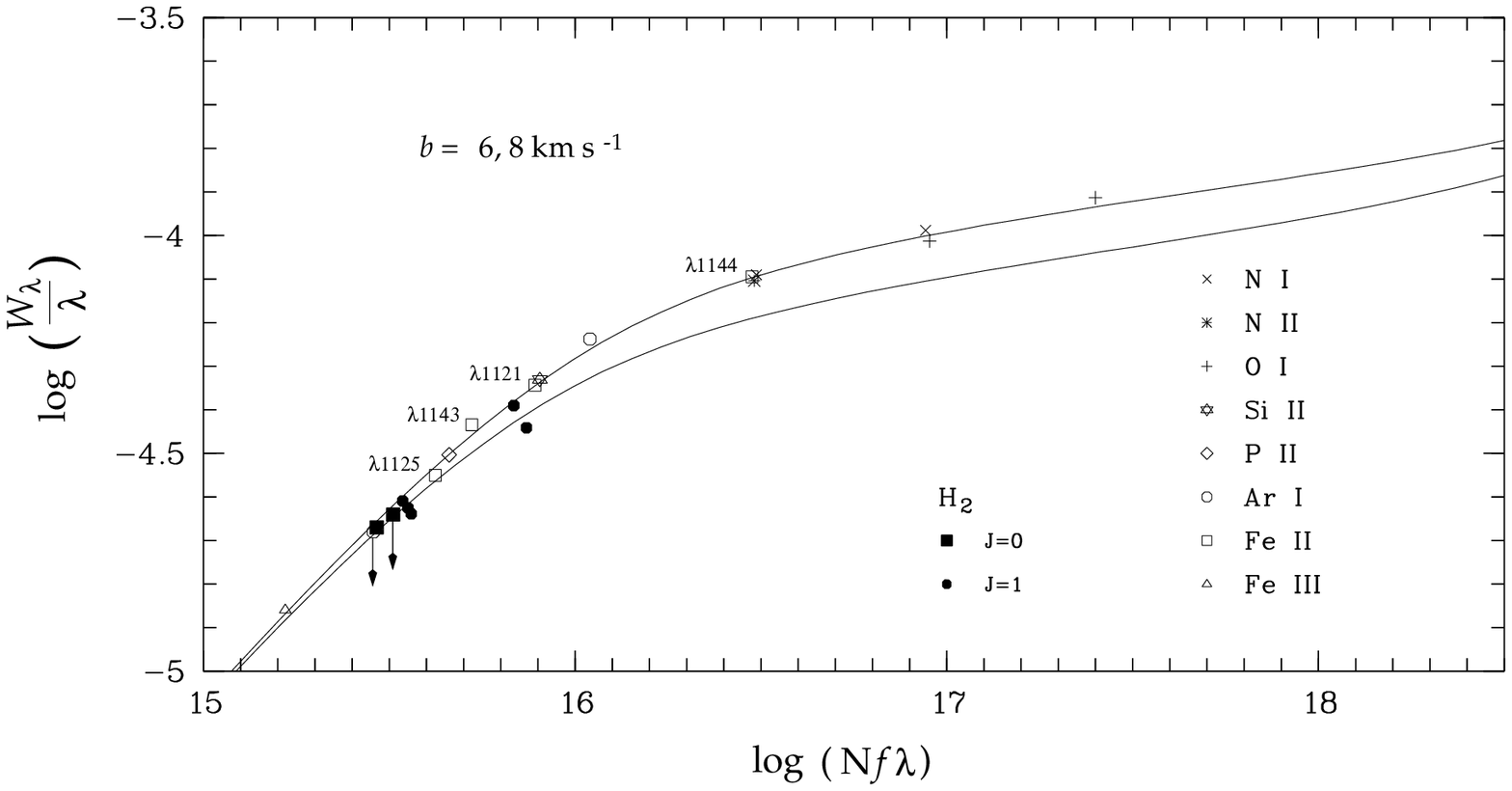}}
\figcaption[pg804fig04.ps]{
Empirical curves of growth for the ion states and molecular hydrogen levels detected in the LLIV Arch toward PG\,0804\,+761.
The ions collectively fit on a curve of growth with $b=8$ km\,s$^{-1}$, defined by four Fe\,{\sc ii} lines. The data points 
for these lines are shown together with their wavelengths. The 
molecular hydrogen lines from $J=0$ and $1$ fit on a curve of growth with $b=6$ km\,s$^{-1}$. The individual species 
are labeled in the lower right corner.
\label{fig5}}

\clearpage
\resizebox{0.85\hsize}{!}{\includegraphics{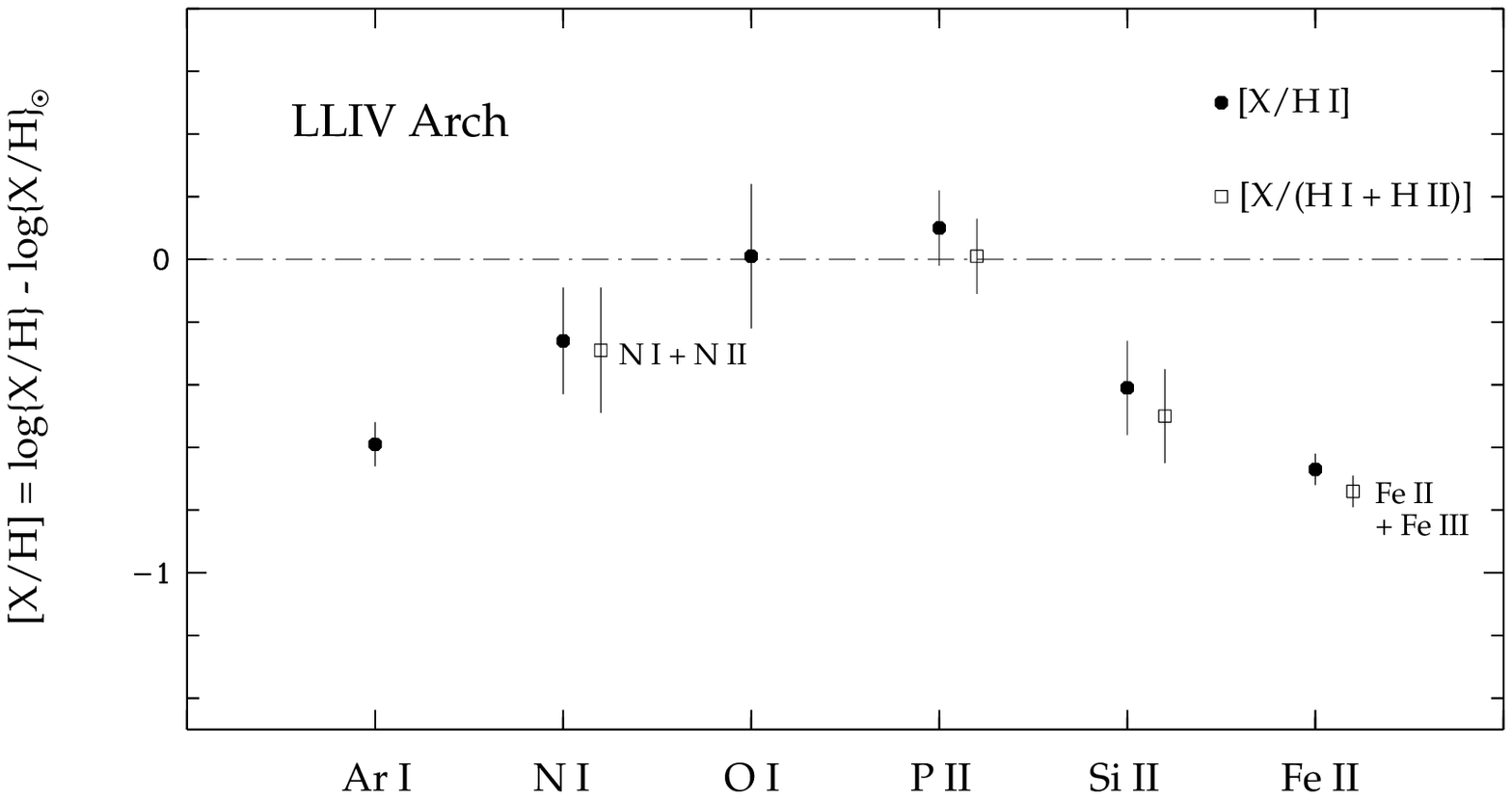}}
\figcaption[pg804fig05.ps]{
Normalized gas-phase abundances of several ionic species associated with gas in the
LLIV Arch at $-55$ km\,s$^{-1}$. Gas phase abundances are referenced to H\,{\sc i} (filled circles) and 
H\,{\sc i}+H\,{\sc ii} (open squares);
they are listed in Table 6. The elements 
are sorted by their condensation temperature (increasing from left to the right, see Savage \& Sembach 1996). 
\label{fig6}}

\clearpage
\resizebox{0.50\hsize}{!}{\includegraphics{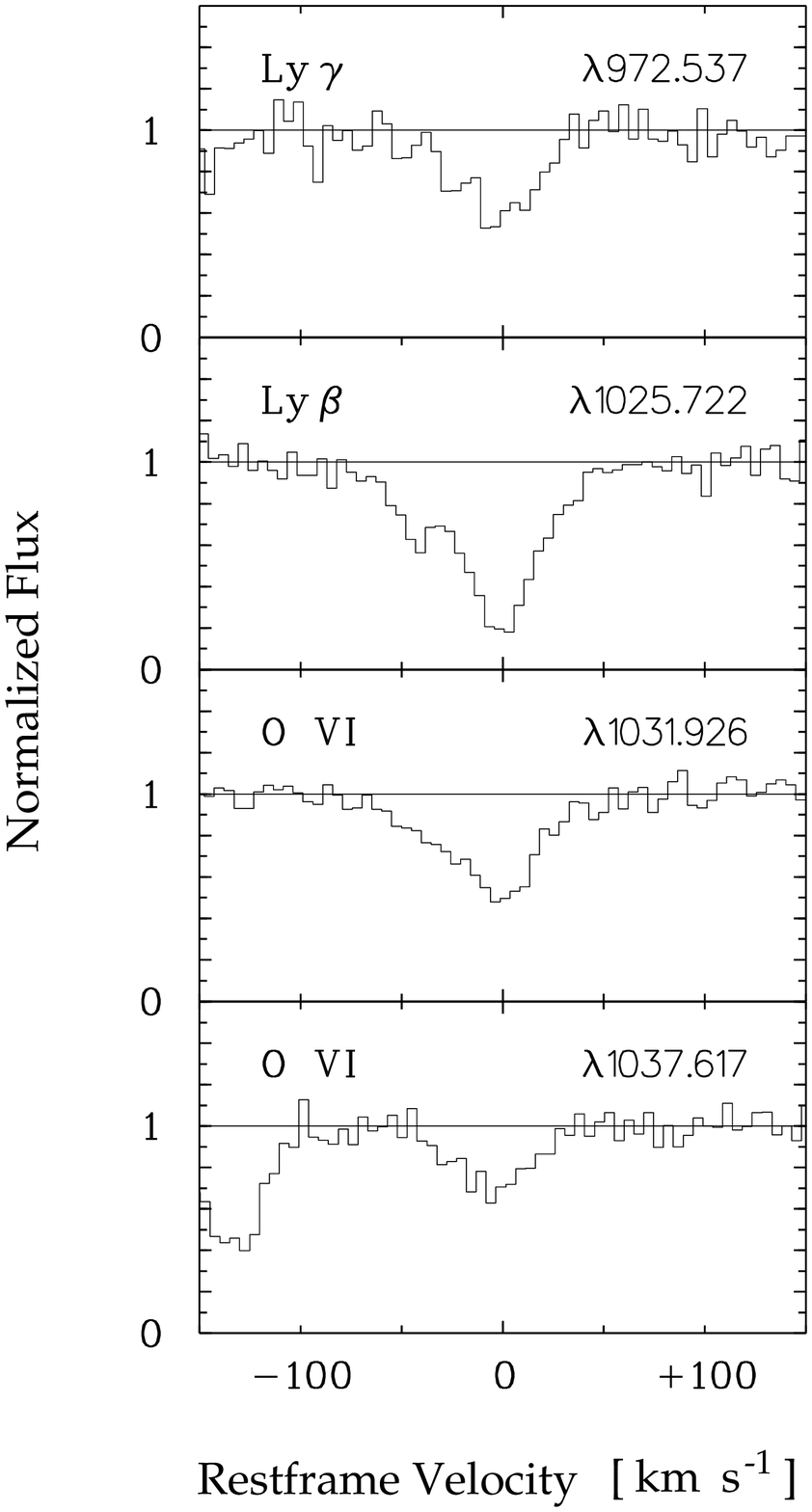}}
\figcaption[pg804fig06.ps]{
Velocity profiles for the absorbing system at $z_{\rm abs} = 0.102$
associated with the background quasar PG\,0804\,+761. The velocity scale
is referenced to the redshift of $z = 0.102$. The associated system is
seen in H\,{\sc i}, C\,{\sc iii}, and O\,{\sc vi} and shows a two-component structure
in the strong lines of H\,{\sc i} $\lambda 1025.722$ and
O\,{\sc vi} $\lambda 1031.926$ with the main component at $0$ km\,s$^{-1}$ and a weaker component
at $\sim -45$ km\,s$^{-1}$. The profile for C\,{\sc iii} is not shown here because it is 
strongly blended by Galactic H$_2$ R(0),2-0 absorption.
\label{fig6}}

\clearpage

\begin{deluxetable}{cccc}
\tabletypesize{\normalsize}
\tablecaption{Log of FUSE Observations}
\tablewidth{0pt}
\tablehead{
\colhead{Target Name} & \colhead{FUSE Dataset Name} & \colhead{Obs. Date} & \colhead{Exp. Time} \\
\colhead{} & \colhead{} & \colhead{} & \colhead{[sec]}
}
\startdata
PG\,0804+761 & P1011903001 & Jan.\,04 2000 & 3464$^{\rm a}$ \\
PG\,0804+761 & P1011903002 & Jan.\,04 2000 & 4730$^{\rm a}$ \\
PG\,0804+761 & P1011903003 & Jan.\,04 2000 & 4449 \\
PG\,0804+761 & P1011903004 & Jan.\,04 2000 & 4291 \\
PG\,0804+761 & P1011903005 & Jan.\,04 2000 & 4191 \\ 
 
\enddata

\tablenotetext{a}{No SiC\,1 channel.}

\end{deluxetable}

\clearpage

\begin{deluxetable}{clcccr}
\tabletypesize{\normalsize}
\tablecaption{H$_2$ Absorption Lines at the Velocity of the LLIV Arch$^{\rm a}$ \label{tbl-1}}
\tablewidth{0pt}
\tablehead{
\colhead{$J$} & \colhead{Transition}   & \colhead{$\lambda_{\rm vac}$\,$^{\rm b}$}
& \colhead{log\,$\lambda f^{\rm b}$}   &
\colhead{$W_{\lambda}$\,$^{\rm c}$} & \colhead{Channel} \\
\colhead{} & \colhead{} & \colhead{[\AA]}   & \colhead{}   &
\colhead{[m\AA]} & \colhead{}
}
\startdata
0 & Lyman  R(0),8-0 & 1001.828 & 1.432 & 21 $\pm$ 9 & LiF\,1A, LiF\,2B  \\
0 & Lyman  R(0),6-0 & 1024.376 & 1.473 & $\le$ 42 & LiF\,1A, LiF\,2B  \\
1 & Lyman  R(1),8-0 & 1002.457 & 1.256 & 36 $\pm$ 9 & LiF\,1A, LiF\,2B  \\
1 & Lyman  P(1),8-0 & 1003.302 & 0.931 & 25 $\pm$ 8 & LiF\,1A, LiF\,2B  \\
1 & Lyman  P(1),5-0 & 1038.158 & 0.956 & 24 $\pm$ 6 & LiF\,1A, LiF\,2B  \\
1 & Lyman  R(1),4-0 & 1049.960 & 1.225 & 43 $\pm$ 6 & LiF\,1A, LiF\,2B  \\
1 & Lyman  R(1),2-0 & 1077.702 & 0.919 & 26 $\pm$ 7 & LiF\,1A \\
2 & Werner Q(2),0-0 & 1010.941 & 1.381 & $\le$ 15 & LiF\,1A, LiF\,2B  \\
2 & Lyman  P(2),5-0 & 1040.368 & 1.017 & $\le$ 21 & LiF\,1A, LiF\,2B  \\
3 & Lyman  R(3),6-0 & 1028.988 & 1.243 & $\le$ 28 & LiF\,1A, LiF\,2B  \\
3 & Lyman  P(3),5-0 & 1043.504 & 1.060 & $\le$ 19 & LiF\,1A, LiF\,2B  \\
3 & Lyman  R(3),4-0 & 1053.975 & 1.137 & $\le$ 29 & LiF\,1A, LiF\,2B  \\

\enddata
\tablenotetext{a}{Equivalent widths are reported for H$_2$ absorption
at the velocity $v_{\rm LSR} \approx -55$ km\,s$^{-1}$ which associates
the absorption with the LLIV Arch.}
\tablenotetext{b}{Vacuum wavelengths and oscillator strengths from Abgrall \& Roueff (1989).}
\tablenotetext{c}{Equivalent widths, $1\sigma$ errors, or upper limits are listed.}
\end{deluxetable}

\clearpage

\begin{deluxetable}{ccc}
\tabletypesize{\normalsize}
\tablecaption{H$_2$ Column Densities in the LLIV Arch \label{tbl-1}}
\tablewidth{0pt}
\tablehead{
\colhead{$J$} & \colhead{$b$} & \colhead{log $N$} \\
\colhead{} & \colhead{[km\,s$^{-1}$]} & \colhead{}
}
\startdata
0 & 6 & 14.04 $\pm$ 0.11\\
1 & 6 & 14.61 $\pm$ 0.13\\
2 & 6 & $\le 14.33$\\
3 & 6 & $\le 14.36$\\
\\
Total &   & $14.71  \pm 0.30$\\

\enddata
\end{deluxetable}

\clearpage

\begin{deluxetable}{lrccr}
\tabletypesize{\normalsize}
\tablecaption{Atomic Absorption Lines in the LLIV Arch \label{tbl-1}}
\tablewidth{0pt}
\tablehead{
\colhead{Species} & \colhead{$\lambda_{\rm vac}$\,$^{\rm a}$}   & \colhead{log\,$\lambda f^{\rm a}$}   &
\colhead{$W_{\lambda}$\,$^{\rm b}$} & \colhead{Channel} \\
\colhead{} & \colhead{[\AA]}   & \colhead{}   &
\colhead{[m\AA]} & \colhead{}
}
\startdata
H\,{\sc i} \dotfill    & 937.804 & 0.864 & 225 $\pm$ 32$^{\rm c}$ & SiC\,1B,  SiC\,2A \\
C\,{\sc ii} \dotfill  & 1036.337 & 2.104 & $\le$ 94     & LiF\,1A, LiF\,2B  \\
N\,{\sc i} \dotfill   &  953.415 & 1.098 & $\le$ 80 & SiC\,2A  \\
                      & 1134.165 & 1.238 & 92  $\pm$ 6 & LiF\,1B, LiF\,2A  \\
                      & 1134.980 & 1.693 & 117 $\pm$ 7 & LiF\,1B, LiF\,2A  \\
N\,{\sc ii} \dotfill  & 1083.990 & 2.097 & 85 $\pm$ 21 & SiC\,2B  \\
O\,{\sc i} \dotfill   &  936.630 & 0.534 &  91 $\pm$ 18 & SiC\,2A  \\
                      & 1039.230 & 0.980 & 127 $\pm$ 30 & LiF\,1A, LiF\,2B  \\
Si\,{\sc ii} \dotfill & 1020.699 & 1.225 &  48 $\pm$ 7 & LiF\,1A, LiF\,2B  \\
P\,{\sc ii} \dotfill  & 1152.818 & 2.451 &  36 $\pm$ 7 & LiF\,1B, LiF\,2A  \\
Ar\,{\sc i} \dotfill   & 1048.220 & 2.440 &  61 $\pm$ 7 & LiF\,1A, LiF\,2B  \\
                       & 1066.660 & 1.857 &  $\le$ 22 & LiF\,1A, LiF\,2B  \\
Fe\,{\sc ii} \dotfill & 1121.975 & 1.512 &  51 $\pm$ 5 & LiF\,1B, LiF\,2A  \\
                      & 1125.447 & 1.244 &  32 $\pm$ 4 & LiF\,1B, LiF\,2A  \\
                      & 1133.665 & 0.728 & $\le$ 18 & LiF\,1B, LiF\,2A  \\
                      & 1143.226 & 1.342 &  42 $\pm$ 5 & LiF\,1B, LiF\,2A  \\
                      & 1144.938 & 2.096 & 92 $\pm$ 6 & LiF\,1B, LiF\,2A  \\
Fe\,{\sc iii} \dotfill & 1122.526 & 2.260 & 16 $\pm$ 5 & LiF\,1B, LiF\,2A \\

\enddata
\tablenotetext{a}{Vacuum wavelengths and oscillator strengths from Morton (in preparation).}
\tablenotetext{b}{Equivalent widths, $1\sigma$ errors, or upper limits are listed.}
\tablenotetext{c}{Including Galactic D\,{\sc i} $\lambda$937.548.}

\end{deluxetable}

\begin{deluxetable}{lccc}
\tabletypesize{\normalsize}
\tablecaption{Atomic Column Densities in the LLIV Arch \label{tbl-1}}
\tablewidth{0pt}
\tablehead{
\colhead{Species} & \colhead{I.P.}   & \colhead{$b$} & \colhead{log $N^{\rm a}$} \\
\colhead{} & \colhead{[eV]} & \colhead{[km\,s$^{-1}$]} & \colhead{}
}
\startdata
H\,{\sc i} & 13.60 & \nodata & 19.54 $\pm 0.01^{\rm b}$ \\
C\,{\sc ii} & 24.38 & 8 & $\le$ 15.20 \\
N\,{\sc i} & 14.53 & 8 & 15.25 $\pm$ 0.17 \\
N\,{\sc ii} & 29.60 & 8 & 14.39 $\pm$ 0.20 \\
O\,{\sc i} & 13.62 & 8 & 16.42 $\pm$ 0.23 \\
Si\,{\sc ii} & 16.35 & 8 & 14.68 $\pm$ 0.15 \\
P\,{\sc ii} & 19.73 & 8 & 13.21 $\pm$ 0.12 \\
Ar\,{\sc i} & 15.76 & 8 & 13.60 $\pm$ 0.07 \\
Fe\,{\sc ii} & 16.16 & 8 & 14.38 $\pm$ 0.05 \\
Fe\,{\sc iii} & 30.65 & 8 & 12.97 $\pm$ 0.10 \\

\enddata
\tablenotetext{a}{The $1\sigma$ column density errors listed for
the metal lines do not include the systematic error associated with
the possibility that the different species do not follow the
curve of growth defined by Fe\,{\sc ii}.}
\tablenotetext{b}{From 21\,cm emission line data with a beam size
of $9\farcm1$ centered on PG\,0804+761.}

\end{deluxetable}

\clearpage

\begin{deluxetable}{lccc}
\tabletypesize{\normalsize}
\tablecaption{Normalized Gas Phase Abundances in the LLIV Arch \label{tbl-1}}
\tablewidth{0pt}
\tablehead{
\colhead{Element} & \colhead{log($X$/H)$_{\odot}$\,$^{\rm a}$}   & \colhead{[$X$/H\,{\sc i}]$^{\rm b}$} 
& \colhead{[$X$/(H\,{\sc i}+H\,{\sc ii})]$^{\rm c}$}\\
\colhead{$X$} & \colhead{+12.00} & \colhead{}\\
}
\startdata
C\,{\sc ii}             & 8.55 & \nodata & \nodata\\
N\,{\sc i}              & 7.97 & $-$0.26 $\pm$ 0.17 & \nodata\\
N\,{\sc i}+N\,{\sc ii}  & 7.97 & \nodata & $-$0.29 $\pm$ 0.20\\
O\,{\sc i}              & 8.87 & $+$0.01 $\pm$ 0.23 & \nodata\\
Si\,{\sc ii}            & 7.55 & $-$0.41 $\pm$ 0.15 & $-$0.50 $\pm$ 0.15\\
P\,{\sc ii}             & 5.57 & $+$0.10 $\pm$ 0.12 & $+$0.01 $\pm$ 0.12\\
Ar\,{\sc i}             & 6.65 & $-$0.59 $\pm$ 0.07 & \nodata\\
Fe\,{\sc ii}            & 7.51 & $-$0.67 $\pm$ 0.05 & \nodata\\
Fe\,{\sc ii}+Fe\,{\sc iii} & 7.51 & \nodata & $-$0.74 $\pm$ 0.05\\

\enddata
\tablenotetext{a}{Anders \& Grevesse (1989); Grevesse \& Noels (1993).}
\tablenotetext{b}{[$X$/H\,{\sc i}] = log($N_X/N_{\rm H\,I})- $log($X$/H)$_{\odot}$.}
\tablenotetext{c}{[$X$/(H\,{\sc i}+H\,{\sc ii})] = log($N_X/N_{\rm H\,I+H\,II})- $log($X$/H)$_{\odot}$.}

\end{deluxetable}

\clearpage

\clearpage
 
\begin{deluxetable}{lcccc}
\tabletypesize{\normalsize}
\tablecaption{Intergalactic Absorption Lines toward PG\,0804+761$^{\rm a}$ \label{tbl-1}}
\tablewidth{0pt}
\tablehead{
\colhead{Species} & \colhead{$\lambda_{\rm obs}$\,$^{\rm a}$}   
& \colhead{$z_{\rm abs}$\,$^{\rm b}$} & \colhead{$W_{\lambda}$\,$^{\rm c}$}
& \colhead{log $N$}\\
\colhead{} & \colhead{[\AA]} & \colhead{} & \colhead{[m\AA]} & \colhead{}\\
}
\startdata
\hline
\multicolumn{5}{c}{Intervening System$^{\rm d}$}\\
\hline
Ly\,$\beta$ & 1044.642 & 0.0185 $\pm$ 0.0004 & 73 $\pm$ 19 & 14.00$^{+0.10}_{-0.13}$\,$^{\rm g}$\\
\hline
\multicolumn{5}{c}{Associated System}\\
\hline
Ly\,$\gamma$ & 1072.050 & 0.1023 $\pm$ 0.0004 & 62 $\pm$ 12 & 14.27$^{+0.07}_{-0.10}$\,$^{\rm g}$\\
Ly\,$\beta$ & 1130.605 & 0.1023 $\pm$ 0.0004 & 172 $\pm$ 11$^{\rm e}$ & 14.37$\pm 0.03$\,$^{\rm g}$\\
C\,{\sc iii} & 1076.994 & 0.1023 $\pm$ 0.0017 & \nodata$^{\rm f}$ & \nodata\\
O\,{\sc vi} & 1137.483 & 0.1023 $\pm$ 0.0004 & 108 $\pm$ 8$^{\rm e}$ & 14.04$^{+0.03}_{-0.05}$\,$^{\rm h}$\\
O\,{\sc vi} & 1143.613 & 0.1022 $\pm$ 0.0004 & 49 $\pm$ 7 & 14.00$^{+0.04}_{-0.05}$\,$^{\rm h}$\\

\enddata
\tablenotetext{a}{The unidentified absorption feature near 1142 \AA\,(see Fig.\,2) is most likely an instrumental
artefact but not an intergalactic absorption line.}
\tablenotetext{b}{Heliocentric wavelengths.}
\tablenotetext{c}{Vacuum wavelengths from Morton (in preparation).}
\tablenotetext{d}{Equivalent widths in the rest frame are listed.}
\tablenotetext{e}{O\,{\sc vi} absorption at $z=0.0185$ is not detected. $3 \sigma$ upper limits
for the two O\,{\sc vi} lines at $\lambda_{\rm rest} = 1031.926$ and $1037.617$ \AA\, are 64 and 52 m\AA,
respectively.}
\tablenotetext{f}{Two-component structure.}
\tablenotetext{g}{Blended by Galactic H$_2$ R(0),2-0.}
\tablenotetext{h}{Column density calculated via $N=1.13 \times 10^{17} W_{\lambda}/f \lambda_{0}^2$,
assuming optically thin absorption.}
\tablenotetext{i}{Column density calculated using apparent-optial-depth method (see Savage \& Sembach 1991).}
\end{deluxetable}

\end{document}